# Nano-topographical changes in latent fingerprint due to degradation over time studied by Atomic force microscopy- option to set a timeline?

First author: **Tereza Svatoňová**, University of Chemistry and Technology, Department of Physics and Measurement, Technická 5,Prague 160 00, Czech Republic

Corresponding and last author: **Anna Fucikova**, Charles University, Faculty of Mathematics and Physic, Department of Chemical Physic and Optics, Ke Karlovu 3, Prague 12116 , Czech Republic

**ABSTRACT**

Latent fingerprints, if present, are crucial in identifying the suspect who was at the crime scene. If there are many latent fingerprints or the suspect is from the same household, crime investigators may have difficulty identifying whose latent fingerprints are time-related to the crime. Here, we report changes in the nanoscale topography of latent fingerprints, which may serve as a timeline and could help estimate when the latent fingerprint was imprinted. On the latent fingerprint of an adolescent, we observed a change in nano-topography over time, specifically the formation of nano-chain structures in space between the imprinted papillary ridges. We consequently compared this observation with the decomposition of the latent fingerprints of a child and adult. We observed a significant difference in the time change in nano-topography of latent fingerprints of a child, adolescent, and young adult. The nano-topographical changes of latent fingerprints were studied by atomic force microscopy over 70 days. In the case of child's and adolescent's latent fingerprints, the first nano-chains were observed already 24 hours after imprinting of the latent fingerprint, and the number of nano-chains increased steadily up to 21 days; then we observed that another organic material covered the nano-chains, and they started slowly deteriorating; nevertheless, the nano-chains were still present on the 70th day.

**INTRODUCTION**

Already at the turn of the 18th and 19th centuries, Eugène-François Vidocq was able to convict perpetrators based on the morphology of the footprint. Forensic techniques are much more advanced in identifying the person, whether the suspect, the victim, or the witness. There are usually several biological traces in crime scenes, such as blood, DNA, hair, handwriting, fibers, footprints, and latent fingerprints. These are very useful, unique, and persistent[1], and serve as evidence in crime scene analysis. In our study, we examined several latent fingerprints imprinted on mica; we were interested in the degradation of transferred substances from the finger skin surface to the place of contact. Many published studies focused on describing the skin surface substance's chemical composition [2–4] and their degradation in time[5,6]. These studies show that skin secretion originates from three sources; the first two are present in the skin, and the third originates from external contaminants (e.g., cosmetics, food remnants, drugs, dust and others). The first source is the epidermis, which forms the upper part of the skin, and the second is the dermis, a layer present below the epidermis. The surface of the epidermis naturally undergoes desquamation and releases about 400 different proteins, of which the major components are keratins 1 and 10 and cathepsin D[4]. In the dermis, eccrine and sebaceous glands are present. These eccrine glands mainly produce proteins, lactic acid, chlorides, amino acids, and other substances in lower quantities. The sebaceous glands produce primarily free fatty acids, wax esters, di-, tri-, and monoglycerides, cholesterol, squalene, and its oxidation products[3].

Generally, the structure of latent fingerprints[7] is studied by a forensic method called dactyloscopy, where the unique pattern of the latent fingerprint serves as the person's identification. DNA can be retrieved from the organic material left on the crime scene[8], even from latent fingerprints [9,10]. In our study, we imprinted the fingerprint on mica, a standard material used in AFM (atomic force microscopy) studies; we chose this substrate since freshly cleaved mica is atomically flat, sterile, and translucent. In the latent fingerprint, we focused on the organic nanostructures on the mica in zones between imprinted material from papillary ridges and studied the deterioration of these substances over time. The imprinted material from papillary ridges in the latent fingerprint cannot be analyzed using AFM since they are higher than the AFM's working height, which is about 10 μm. Also, these zones did not show an interesting change in time other than occasional shrinkage of the imprinted papillary line width. In this work we observed morphological and chemical differences in latent fingerprints of people who belong to different age categories. The work of Mong et al.[11] shows a representation of chemical components produced by sweat glands in the skin of people of all ages, and the data shows that the substances produced by sebaceous glands are primarily present in adult latent fingerprints. These include, for example, the less volatile longer chain fatty acid esters. For very young individuals, aqueous salt solutions prevailed in the latent fingerprints, and in contrast, cholesterol predominated in the latent fingerprints of adolescents[11]. On the one hand, the latent fingerprints of children contain free fatty acids, whereas in the latent fingerprints of an adult, there are less volatile long-chain fatty acid esters[12,13]. According to the study[14], all children's latent fingerprint imprints were mainly composed of three groups of substances: esters, secondary amides, and acid salts. The latent fingerprint of a child or a person of prepubescent age primarily consists of sodium lactate[14].

In the work of D.K. Williams et al.[14], they investigated the degradation of substances originating from the skin in the latent fingerprints; they found that the number of substances containing double bonds (squalene and others) decreased significantly within a month in the imprinting. The number of saturated molecules, e.g., wax esters and cholesterol, also decreased in time, but at a much slower rate. At the same time, the amount of low molecular weight saturated acids formed from the oxidized products of squalene and some fatty acids increased.

Currently, one can only distinguish significantly old latent fingerprints from newer ones on the crime scene due to dust or other impurities present on the old latent fingerprints. Sometimes, one can identify which fingerprint was imprinted first due to its overlap with another fingerprint. The tool for the precise age of latent fingerprint is still missing; based on our observation, we hope our discovery can help develop such a tool.

In our work presented here, we focused on monitoring the degradation of fingerprint imprints in different age categories over time and comparing them to the degradation of specific substances present on the skin, namely squalene and lactic acid. We also studied latent fingerprints after washing hands with soap and the natural restoration of skin secretion. The nano-topography changes during the aging of latent fingerprints and comparison samples were examined using AFM over up to 70 days.

## MATERIALS AND METHODS

### Materials

Chemicals: Squalene (100% pure Squalene) was purchased from Saloos produced by Mica and Harasta s.r.o. Czech Republic, sodium lactate (pure in water) was purchased from Likochem, Czech Republic. Sodium lactate was diluted with deionized water (prepared by Aqual 29, from company Aqual s.r.o , Czech Republic).  As standard soap, we used generic liquid soap Mitia from TOMIL s.r.o. Czech Republic.  V-1 grade Mica was purchased from 2Spi.com USA. A transparent Confocal-matrix® adhesive from company ML chemica, Czech Republic, was used to glue the mica to a standard microscopic slide.

### Preparation of samples

Here presented results consist of latent fingerprints of the index finger of a child (girl, 8 years), an adolescent (woman, 21 years), and an adult (man, 54 years) to reflect people of different ages. This selection of latent fingerprints was based on our previous experiments, in which we studied the latent fingerprints of the index fingers of five young males (18-24 years) over two months. At least three zones have been studied with atomic force microscopy from each latent fingerprint. The latent fingerprints were imprinted randomly during a usual workday at least one hour after washing hands with soap and water on a freshly cleaved mica. The mica was glued with Confocal-matrix® adhesive to a standard microscopic slide. Just before the imprinting of the index finger to a mica, its upper, potentially dirty layer was removed by duct tape; with this procedure, we obtained an atomically flat, clean, sterile, and negatively charged mica surface.  The micas with latent fingerprints samples were stored at room temperature (stable 20 °C, air-conditioned room) in standard humidity (ranging from 40-60 %) and kept in the dark except for AFM measurement (2-4 hours per sample per measurement, light coming from lamp TH4-200 from Olympus via inverted microscope).

In one case (woman, 21 years old) we decided to investigate the effect of washing hands with soap on the latent fingerprint degradation. The latent fingerprints (of a woman, 21 years old) were imprinted immediately after washing the hands with soap (from Mitia), 20 min after washing, and 30 min after washing. There was no contact of the finger with any object or part of the body between the hand washing and latent fingerprint imprinting.

As a reference sample, we decided to study the degradation of two organic substances, squalene and sodium lactate, which naturally occur on the skin and are present in different quantities in different age groups according to the literature[14,15]. In the case of squalene, 1 μl of it was applied on a sterile nitrile glove, and then this material was imprinted on a freshly cleaved mica. In the case of sodium lactate, we prepared an aqueous solution in a ratio of 1:5 and 2 μl was drop-cast on a freshly cleaved mica.

**The atomic force microscope setup**

For the measurement of samples nano-topography, we used atomic force microscope Nanowizard 3 from the JPK Instruments company (now Bruker). The device was controlled via JPK NanoWizard Control software from the JPK company, and then the images were processed using JPK Data Processing software from the same company. All AFM images were taken through AC mode with an ACTA cantilever from the App Nano company with a nominal frequency of 300 kHz, stiffness k = 37 N/m, and sharpness of the tip of 6 nm. The AFM is placed on an inverted microscope Olympus IX73, where the light is provided by a halogen lamp (Olympus TH4-200) and is guided via 10x UPlanFL N 0.80 NA objective from the same company. All AFM images were taken in a resolution of 256 x 256 px. The attached images are in the size of 10 x 10 μm, and 4 x 4 μm respectively. The scan rate was set around 0.2 Hz.

**RESULTS AND DISCUSSION**

**The investigation of latent fingerprint degradation**

The first set of AFM images of latent fingerprints was taken a few hours after imprinting, and the last AFM images were taken about 70 days after the latent fingerprint was created. In each AFM measurement, the precise measured location in the latent fingerprint was found thanks to the marks made by a marker on the other side of the mica and colocalization procedure (for detailed information, see support information S1). At least three zones have been observed from each latent fingerprint. Finding the exact location and measurement of a single AFM image takes around 1 hour, which means it took 3 to 4 hours of measurement per one latent fingerprint at a given time point. Therefore, latent fingerprints from each individual in this study were created on a different day so that we could measure the sample degradation within the same time spacing. The first fingerprint measurement (named here as 1st day) was done at a maximum of 4 hours after latent fingerprint creation.

In Fig.1, we present one zone from the latent fingerprint of a child (an 8-year-old girl); AFM images were taken on the 1st, 2nd, 7th, 20th, 37th, and 70th day from the imprinting of a fingerprint (enlarged image is given in supplementary info Fig. S2-1).

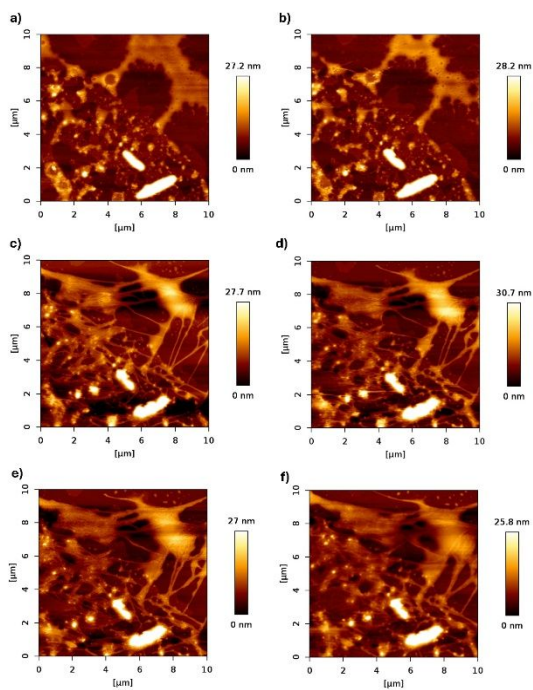

Figure 1: AFM images of the latent fingerprint of an 8-year-old girl taken on a) the first day, b) the second day, c) the seventh day, d) the twentieth day, e) thirty-seven day, and f) seventy days from imprinting (detailed image is given in supplementary info in figure S2-1).

Roughly after 24 hours (named here as $2^{nd}$ day of measurement), we observed the first formation of nano-chained structures, as we can see in Fig. 1a) in the upper right corner (a detailed image with an arrow pointing to the nano-chain is provided in supp. info figure S2-1b). The number of nano-chains in the AFM image increased significantly on the $7^{th}$ day (Fig1 c.) compared to the AFM measurement done on the first day. The nano-chains are about 4 nm in height and relatively long (up to 2-3μm in Fig. 1c). These nano-chains are present between droplets of organic material originating from the imprinting of papillary ridges in the latent fingerprint. On the 21st day after the imprinting Fig. 1d, we observed that the nano-chains also formed within big puddles of the biological material, whereas the nano-chains, which were single-standing, now appear thicker. On the following measuring days (Fig. 1d) $20^{th}$ day, 1e) $37^{th}$ day, 1f) $70^{th}$ day), we observed that the nano-chains were getting covered by another material, which caused their thickening and formation of clusters of organic matter, but despite that the nano-chains have been observable still on the last day of measurement (Fig. 1f)). The same situation was observed on the other two zones within the same latent fingerprint.

Since the most dramatic change in the AFM images of latent fingerprints and the appearance of the nano-chains happens between the $1^{st}$ and $7^{th}$ day, we present comparative images of fingerprints of a child (8-year-old girl), adolescent (21-year-old woman) and adult (54-year-old man) at these times in Fig. 2. Detailed time study with AFM images taken on $1^{st}$, $2^{nd}$, $7^{th}$, $20^{th}$, $37^{th}$ and $70^{th}$ day are shown in supporting information Figures S2-1 for 8-year-old girl, S2-2 for 21-year-old woman, S2-3 for 54-year-old man.

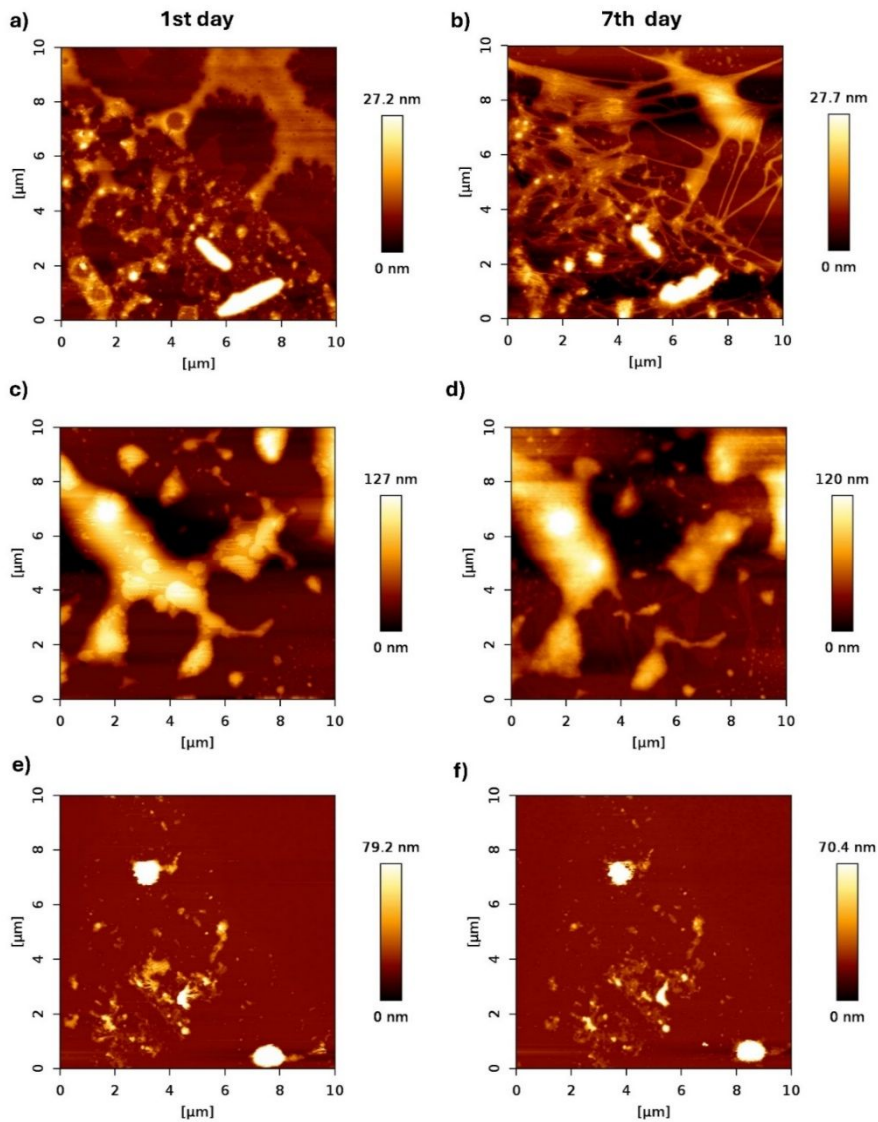

Figure 2: AFM images of fingerprint imprint taken within a maximum of 4 hours after imprinting -1$^{St}$ day (a, c, e) and seven days after imprinting (b, d, f) of an 8-year-old female child (a, b), of a 21-year-old young woman (c, d) and a 54-year-old man (e, f).

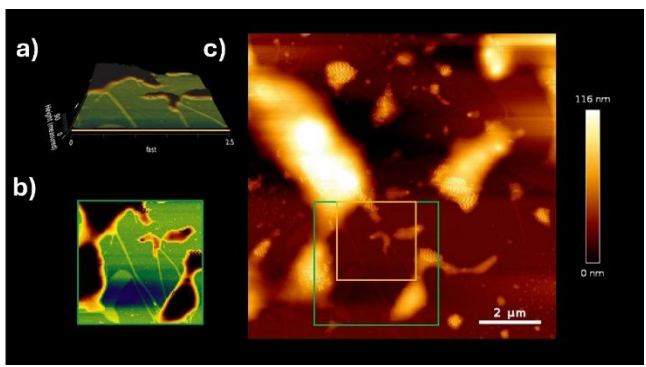

Figure 3: AFM images of fingerprint imprint of 21-year-old female taken on 2$^{nd}$ day from imprinting with a different colour scheme to visualize the individual nano chain structures. a) the 3D image from the yellow square from the big image c); b) the enlarged zone from the green square in c); c) standard 10x10µm AFM image identical to S2b).

The same nano-chained structures were observed on the second day (S2-2b marked with yellow arrow) in the case of the latent fingerprint of a 21-year-old woman; the number of these nanostructures also increased in time, as we can see in (Fig. 2 c), d)). Since the contrast of the nanostructures is not always clearly visible in the AFM image due to the high aspect ratio of organic residues originating from the finger's papillary lines imprinting in the latent fingerprint, we are also presenting them in different colour schemes in Fig. 3. The highest concentration of these clearly visible nano-chains was observed in latent fingerprints around 7-20$^{th}$ day from imprinting in both the imprints of a child and adolescent (see supp. info figures S2-1d), S2-2d)). After this time, the nano-chains have been covered by another organic matter, but they are still visible at day 70 (see supp. info figures S2-1f), S2-2f)).

We did not observe any formation of nano-sized chains in the case of the latent fingerprint of a 54-year-old man as shown in (Fig. 2 e),f) and supp info figure S2-3, respectively).

In our preliminary experiments, before this study, we studied the latent fingerprints of five young males (18-24 years) over two months. In all cases, we observed very similar behaviour of latent fingerprint degradation and the appearance of nano-chains at similar times, with similar features as a look or nano-chain heights, as in the case of the 21-year-old female, whose AFM images of latent fingerprints are presented in this article.

From the observation of the different amounts of nano-chains in the child and several adolescents' latent fingerprints and the lack of such nano-chains being present in the latent fingerprints of a 54-year-old male, we looked closely at what molecules are present in the latent fingerprints of young people and are practically not present in over 40-50-year-old humans. According to article[15], in which the authors claim that the amount of squalene molecules decreases in time, whereas the amount of squalene oxidation products is increasing, we selected to investigate squalene as a possible source of observed nano-chained nanostructures. Another study[14] claims that sodium lactate predominates in the child's latent fingerprint, and its amount decreases in older adults' latent fingerprints.

Based on the literature and our AFM observation of latent fingerprints, we presumed that with high probability, either squalene or lactic acid might be responsible or contribute to the formation of observed nano-chains. We purchased them in the purest available form and drop-cast them on mica as described in the materials and methods. We did not observe any formation of nano-chains in the case of the squalene molecule, as we can see in supp. info Fig. S3-1. On the other hand, in the case of the pure lactic acid molecule study, identical nano-chains to the ones observed in our fingerprint imprint study have already been observed on the 4$^{th}$ day (Fig 4 a)) from drop casting and on the 7$^{th}$ day (Fig 4 b)), the sample was quite fully covered in nano-chains. The delay in the formation of nano-chains in the case of pure lactic acid molecules is probably caused by the low concentration of these molecules in the sample and by the fact that we diluted the molecule significantly with water. Nevertheless, in all the AFM-studied latent fingerprints of the child, adolescents, and pure lactic acid samples, the nano-chains have had an identical height of 3,5 nm ± 0,6 nm and the same nano-topography.

This observation provides very supportive evidence that the molecule that forms or contributes to the formation of the nano-chains observed here in the latent fingerprints AFM images of younger people can very likely be lactic acid. The fact that the amount of lactic acid in latent fingerprints lowers significantly in older adults, as described in the study[14], could explain why we did not observe any formation of nano-chains in the AFM image of the here presented latent fingerprint of the 54-year-old male.

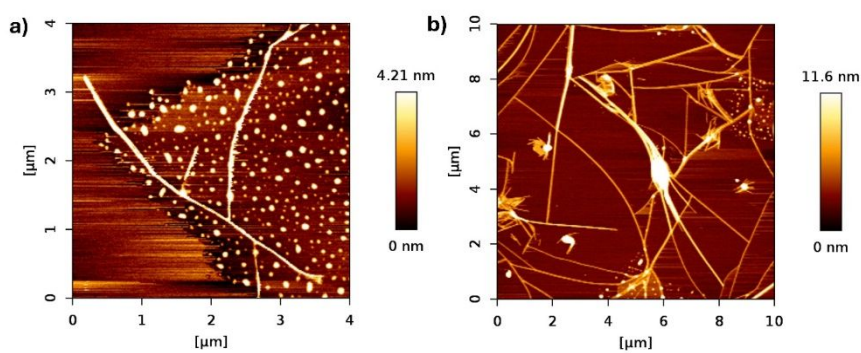

Figure 4: AFM images of pure lactate molecule a) 4 days after drop casting and b) 7 days after drop casting.

**The skin surface convalescence after hand washing with soap**

For practical use, we decided to do a short experiment exploring how washing hands affects the degradation of the latent fingerprint and whether we will observe similar nano-chain structures in latent fingerprint after hand washing. In this case, we selected a 21-year-old female who washed her hands with standard liquid soap (see Methods and Materials), and for a half-hour, the young adult did not touch any objects or other parts of the skin to avoid cross-contamination of the finger. After this time, she imprinted her latent fingerprint on freshly cleaved mica in an identical procedure to that used for the above latent fingerprints. As shown in Fig. 5, the typical nano-chains with identical height are already visible in the AFM images of latent fingerprints after roughly 4 days from imprinting. We can see that for the recovery of the skin secretion and the formation of nano-chains (see Fig. 5b) in the latent fingerprint, it is enough to wait 30 minutes after washing before the latent fingerprint imprinting.

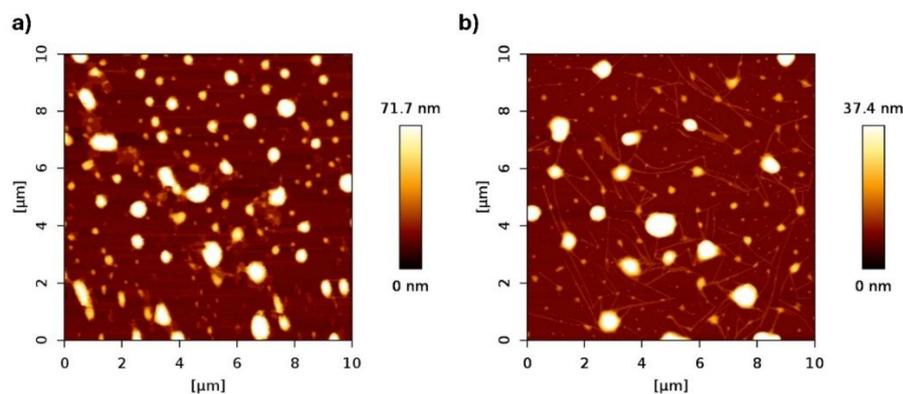

Figure 5: AFM images of latent fingerprint of 21-year-old female 30 minutes after washing with soap a) fresh imprint after 30 min after washing, b) the same spot 3,5 days later.

**CONCLUSION**

In the latent fingerprints of children and adolescents, we observed the formation of nano-chains whose number increases up to roughly 20 days from imprinting. Then, we observed the slow covering of the nano-chains with another biological matter. We can clearly distinguish an old from a new latent fingerprint. If we compare two AFM images of latent fingerprints taken at different times, we can estimate how much time has passed from the imprinting of the latent fingerprint based on the number of nano-chains and their look.

The number of nano-chains and the general appearance of the AFM image of the latent fingerprint also enable us to estimate whether it belongs to a child, adolescent, or adult. There is quite strong support that the observed nano-chains are related to the presence of lactic acid in the latent fingerprint since we observed the appearance of nano-chains in pure lactic acid degradation experiment (Fig. 4) and nano-chains in younger people's latent fingerprints, which is in agreement with the study[14].

The AFM images of 54-year-old male latent fingerprints show no significant change in the nanostructure topography over time. This might be explained by the fact that lactic acid presence in the skin disappears for people over 50. In our opinion, this limitation is not a significant problem since 80-90% of crimes are committed by people younger than 30. Also, it is quite rare that a person over age of 50 is the first crime committer[16–18].

We are fully aware that the real application of our method, determining the age of latent fingerprints by measuring AFM images in situ on the crime scene, is time-consuming and unpractical. But we hope that our observations can lead to the development of, for example, a fluorescent chemistry kit based on the detection of nano-chained structures, which are highly probably connected to the presence of lactic acid in the latent fingerprint, that could be used in real crime scenes.

Acknowledgment:

The authors declare no competing financial interests. This work was supported by UNCE/SCI/010 provided by Charles University.